# In vivo Cancer Marker in Mice

I.I.Trifonova
National Oncological Center,
6 Plovdivsko Pole Str., Sofia 1756,
Bulgaria

S.Z.Stefanov*
NEK - National Dispatching Center,
8 Triaditsa Str., Sofia 1040,
Bulgaria

October 28, 2004

**Abstract**

This work introduces an in vivo cancer marker in mice from the altitude of the well, determined by this cancer. The duration of development and the altitude of the cancer well are obtained from the synchronization of the mechanics and the kinetics of the processes of normal and anomalous transport under tumor mass growth. The mechanics and the kinetics of these two transports are found from the cancer fugacity. At that, it is assumed that these two transports are in a fractal distribution network. The spectrum of this network is also obtained from the fugacity of the cancer under consideration. The cancer fugacity is defined as a photograph of the rate of the scaling exponent under tumor mass growth.

___________________________
e-mail: szstefanov@ndc.bg

# 1. Introduction

The in vivo growth of tumors can be described by a law of growth of the tumor mass [6]. The consideration of the tumor mass under this law of growth corresponds to the development of cancer in a tissue which is a physical soft matter. The classical laws of tumor growth are laws of growth of the tumor volume [8]. They disregard the tissue in which the cancer develops. That is why the law of the in vivo growth of tumors is a more adequate representation of the actual development of tumors.

We shall assume that the tumor is a well of the energy needed for its in vivo growth. Here the well of energy will be understood in accordance with Aubin and Lesne [1] as duration in time and as altitude.

The aim of this work is to find a marker of the in vivo cancer in mice from the altitude of the well, determined by this cancer.

# 2. Fugacity of the in vivo Cancer

The in vivo growth of tumors can be described [6] by West's universal law of growth

$$dm/dt = vm^{p(t)} (1-m/M)^{1-p(t)} \qquad (1)$$

In (1) m is the actual mass, dm/dt is the rate of growth, M is the asymptotic value of m(t), v characterizes the cell-line metabolic rate, p(t) is the scaling exponent of growth.

The scaling exponent p(t) is obtained in [6] for three lines of tumor cells. It has been found out that, in these cases, the scaling exponent first decreases, and then increases. It has been suggested [6] that this behavior of the scaling exponent is due to the change of the tumor's dominating nutrient-supply mechanism from passive diffusion to active perfusion (angiogenesis). The decreasing scaling exponent will be denoted by $p_1(t)$, and the increasing one - by $p_2(t)$.

The scaling exponent rates a and b are obtained from the data taken from the three figures in [6], under the following regression between them and time

$$p_1(t) \sim \exp(at + c) \qquad (2)$$
$$p_2(t) \sim \exp(bt + d)$$

The values of (a, b) for the three cell lines are

$$(a_1, b_1) = (4.5143*10^{-3}, -1.8692*10^{-2}) \qquad (3)$$
$$(a_2, b_2) = (-6.0405*10^{-2}, 7.4514*10^{-3})$$
$$(a_3, b_3) = (-6.1118*10^{-3}, 1.0844*10^{-3})$$

The scaling exponent rate will be replaced by its photograph. The photographic scaling down is ten times in space, and it is $k\pi$, k - an integer, in time [5].

The photograph of the scaling exponent rate, $\alpha$, under diffusion, is obtained from a and b, and it is

$$\alpha_1 = a_1 10^3/\pi, \; \alpha_2 = b_2 10^3/2\pi, \; \alpha_3 = -a_3 10^3/\pi \qquad (4)$$

The photograph of the scaling exponent rate, $\beta$, under perfusion, is obtained from a and b, and it is

$$\beta_1 = -b_1 10^2/\pi, \; \beta_2 = -a_2 10^2/4\pi, \; \beta_3 = -b_3 10^3/\pi \qquad (5)$$

The photographed rates of the scaling exponent show the fugacity of the in vivo cancer development.

DEFINITION 1. The fugacities of the in vivo cancer development in mice are the photographed rates (4), (5) of the scaling exponent p(t).

## 3. In vivo Cancer Marker

Let the photographed rates of the scaling exponent set the spectrum of the fractal distribution network from [6]. The spectrum of these fractals has an angular speed $\theta$ [7]

$$\theta = (((2+\alpha)^2 - 4\alpha(1-\cos(2\pi\beta)/2))/(4\alpha))^{1/2} \qquad (6)$$

The in vivo cancer development will be regarded as a system with normal and anomalous transport in the fractal distribution network. The predominant transport in the fractal distribution network will be considered to be normal under passive diffusion, and anomalous - under active perfusion.

The mechanics of this system with normal and anomalous transport gives the moment of termination of the conflict between these two processes.

The kinetics of this system with normal and anomalous transport gives the moment of domination of the anomalous transport.

The coincidence of the two moments is synchronization of the mechanics and kinetics of the above system with transport. Let the common moment of the mechanics and kinetics be denoted by t*.

We shall assume that this moment t* is the critical moment at which the spectral expansion of a non-self-adjoint harmonic oscillator is norm convergent for t>t* and divergent for 0≤t<t*.

From (6) and from the results in [3] it follows that

$$t^* \sim \exp(tg(\theta^*))/2, \; \theta^* = \pi\theta/2 \; \text{ or } \; \theta^* = 3\theta/2 \qquad (7)$$

The values of t* for the three cell lines are

$$t_1^* = 10.97, \; t_2^* = 11.24, \; t_3^* = 12.88 \qquad (8)$$

In (8) $t_1^*$ and $t_2^*$ are obtained for $\theta^* = \pi\theta/2$ in (7), and $t_3^*$ is obtained for $\theta^* = 3\theta/2$ in (7).

The conflict between the normal and the anomalous transport is a game of pursuit. The terminal closeness of the two processes is [2]

$$\varepsilon = (\Gamma(\alpha)/\Gamma(\beta))^{(\beta/\alpha-\beta)} / \Gamma(\beta+1) - (\Gamma(\alpha)/\Gamma(\beta))^{(1-\beta)} / \Gamma(\alpha+1) \quad (9)$$

In (9) $\Gamma(\cdot)$ is the gamma function.

The values of $\varepsilon$ for the three cell lines are

$$\varepsilon_1 = 0.5974, \quad \varepsilon_2 = 0.5511, \quad \varepsilon_3 = 1.0417 \quad (10)$$

The mean of the diffusion for the one dimensional space kinetics of a system with normal and anomalous transport, under domination of the anomalous transport, is [9]

$$m_k \sim (2/\pi)((t^*/2)^{1/2} \Gamma(1-1/\alpha) + \quad (11)$$
$$(t^*/2)^{(1-1/(2-\beta))} \Gamma(1+1/(2-\beta)) + \varepsilon)$$

The mean $m_k$ from (11) characterizes the anomalous transport under in vivo cancer. That is why $m_k$ from (11) sets the following cost of in vivo cancer dissemination

$$c = m_k/2 \quad \text{or} \quad c = m_k/2 - 10 \quad (12)$$

The values of c for the three cell lines are

$$c_1 = 3.73, \quad c_2 = -1.2, \quad c_3 = 2.44 \quad (13)$$

In (13) $c_1$ and $c_3$ are obtained from (12) for $c = m_k/2$, and $c_2$ is obtained from (12) for $c = m_k/2 - 10$.

DEFINITION 2. The in vivo cancer in mice sets a well of energy with duration $t^*$ and altitude c.

The three wells of energy, for the three cell lines, are the following pairs w

$$w_1 = (t_1^*, c_1), \quad w_2 = (t_2^*, c_2), \quad w_3 = (t_3^*, c_3) \quad (14)$$

The zeros of the weight system of the colored Jones function for the "Figure Eight" knot are [4]

$$\lambda_1 = 1.45, \quad \lambda_2 = -2.00, \quad \lambda_3 = -3.45 \quad (15)$$

The values of c from (13) are close to these zeros. The in vivo cancer marker will be defined through this correspondence.

DEFINITION 3. The in vivo cancer marker in mice, $m^*$, is

$$m^* = [c/2] + 2 \quad (16)$$

Here $[\cdot]$ is a rounding of $(\cdot)$.

The marker $m^*$ has the following values

$$m = \begin{cases} 4, \text{ for the first cell line} \\ 1, \text{ for the second cell line} \\ 3, \text{ for the third cell line} \end{cases} \quad (17)$$

# 4. Conclusion

The main result of this work is the defined in vivo cancer marker in mice. For this purpose the following have been obtained:
1) the cancer fugacity;
2) the duration of the cancer development;
3) the altitude of the cancer well of energy.

The marker has been introduced from the altitude of the cancer well of energy.

Whether the above marker is valid for all cell lines of in vivo cancer in mice will be the subject of investigation in a next work. Also in a next work, a check will be made of the validity of the proposed in vivo cancer marker in humans.